\documentstyle[12pt]{article}

\begin{document}
\input{psfig.sty}
\begin{flushright}
\baselineskip=12pt
MADPH-99-1148 \\
\end{flushright}

\begin{center}
\vglue 1.5cm
{\Large\bf Non-Compact $AdS_5$ Universe with Parallel Positive Tension 3-Branes\\}
\vglue 2.0cm
{\Large Tianjun Li}
\vglue 1cm
\begin{flushleft}
Department of Physics, University of Wisconsin, Madison, WI 53706,  
U.  S.  A.
\end{flushleft}
\end{center}

\vglue 1.5cm
\begin{abstract}
We construct the general model with only parallel positive tension 
3-branes on $M^4 \times R^1$ and $M^4 \times R^1/Z_2$. The 
cosmology constant is sectional constant along the fifth dimension. 
In this general scenario, the 5-dimensional GUT scale on each brane 
can be indentified as the 5-dimensional Planck scale, but, the 
4-dimensional Planck scale is generated from the low 4-dimensional 
GUT scale exponentially in our world. We also give two simple models 
to show explicitly how to solve the gauge hierarchy problem.

\end{abstract}

\vspace{0.5cm}
\begin{flushleft}
\baselineskip=12pt
November 1999\\
\end{flushleft}
\newpage
\setcounter{page}{1}
\pagestyle{plain}
\baselineskip=14pt

\section{Introduction}

Experiments at LEP and Tevatron have given the strong
support to the Standard Model of the Strong and electroweak 
interactions. However, the Standard Model has some unattractive features which
may imply the new physics beyond the Standard Model. 
One of these problems is that the gauge forces and the gravitational force
are not unified. Another is
the gauge hierarchy problem between the weak scale and the 4-dimensional
Planck scale. Previously, two solutions to the
gauge hierarchy problem have
been proposed: one is the idea of the technicolor and  compositeness
which lacks calculability, and the other is the idea
of supersymmetry.    

More than one year ago, it was suggested that the large 
compactified extra dimensions may also be the solution to the
gauge hierarchy problem~\cite{AADD}. Moreover,
half year ago, Randall and Sundrum~\cite{LRRS} proposed another scenario
that the extra dimension is an orbifold,
and  the size of the extra dimension is not large
but the 4-dimensional mass scale in the standard model is
suppressed by an exponential factor from 5-dimensional mass
scale. In addition,  they suggested that
the fifth dimension might be infinity~\cite{LRRSN}, and there may exist only one
brane with positive tension at the origin, but, there  exists the  gauge
hierarchy problem.  The remarkable aspect of
the second scenario is that it gives rise to a localized graviton field.
Combining those results, Lykken and Randall obtained the following
physical picture~\cite{JLLR}: the graviton is localized on Planck brane, we
live on a brane separated from the Planck brane about 30 Planck lengths 
along the fifth dimension. 
 On our brane, the mass scale
in the Standard Model is suppressed
exponentially, which gives the low energy scale. Furthermore,
we generalized those scenarios and obtained the scenario with the following
property~\cite{LTJ}: the 4-dimensional Planck scale is generated from the low 5-dimensional
Planck scale by an exponential hierarchy, and the mass scale in the
Standard model, which is contained in the observable brane, is not rescaled.
In short, recently, this kind of compactification or similar idea has attracted
a lot of attentions~[5-27].
By the way, supergravity domain walls was discussed previously in the 4-dimensional
space-time~\cite{MCHS}. 

In the model building, almost all the models~\cite{LRRS, JLLR, IO, LTJ, KRDDG, ODA, HSTT}
which can solve the gauge hierarchy problem
 contain the branes with the negative brane tension except Lykken-Randall model 
 and Oda model~\footnote{
I do not consider the models with brane intersections/junctions here
~\cite{HDDK, CCYS}, and many brane junctions will be considered
elsewhere~\cite{LTJII}.}. However,
mathematically speaking, Lykken-Randall model~\cite{JLLR}
 may not be strict because they neglected the effect from the
observable brane tension which they assumed might be very small,
and in Oda model~\cite{ODA},
 the transformation is not 1-1 and may not be topological invariant.
We want to consider the model with only positive tension branes, because, as we know, 
an object with negative tension can not be stable, although we might not
need to worry about it if the anti-brane is orientable~\footnote{ I would like to thank
J. X. Lu for pointing out this to me.}. In addition,
there are positive energy objects, namely D-branes and NS-branes,
that are well understood  and on which gauge fields and matter fields can be 
localized so that the Standard Model fields can be placed there.

Before our discussion, we would like to explain our  
assumption and notation which are similar to those in~\cite{LTJ}. 
 We assume that all the gauge forces are unified on each 3-brane
 if there exist gauge forces. The 5-dimensional GUT scale
 on i-th 3-brane  $M5_{GUT}^{(i)}$  and the 5-dimensional Planck scale $M_X$ are defined as
 the GUT scale and Planck scale in the 5-dimensional fundamental
 metric, respectively. 
 The 4-dimensional GUT scale on i-th 3-brane $M_{GUT}^{(i)}$ and the 4-dimenisonal
 Planck scale $M_{pl}$ are defined as the GUT scale and Planck scale in the
 4-dimensional Minkowski Metric ($\eta_{\mu \nu}$). In order to
 avoid the gauge hierarchy problem between the weak scale and
 the 4-dimensional GUT scale $M_{GUT}$ on the observable brane which includes
 our world, we assume the low energy unification~\footnote{We 
 will not explain why $M_{GUT}$ can be low energy scale here,
but, it is possible if one considers additional particles which 
change the RGE running. Of course, proton decay might be the
problem, but we do not disscuss this here. These subjects were discussed in ~\cite{KRDDG, 
GUTP}.}. 
The key ansatz is that the 5-dimensional GUT scale on each brane 
 is equal to the 5-dimensional Planck scale.

In this paper, we construct the general model with only parallel positive tension 3-branes. 
If the fifth dimension is compact, the sum of the brane tensions is zero which is
the topological invariant of this kind of models~\cite{LTJI}.
 Therefore, we have to go to
the non-compact fifth dimension case. From differential topology/manifold, up to
diffeomorphic, there is only one connected non-compact 1-dimensional manifold: $R^1$,
and there is only one connected non-compact 1-dimensional manifold with boundary:
$H_1$ or $R^1/Z_2$ ( the equivalence class is $y \sim -y$ ). 
Therefore, the space-time we consider is $M^4\times R^1$ and 
$M^4\times R^1/Z_2$ where $M^4$ is the 4-dimensional Minkowski space-time. The parallel 
positive 
tension 3-branes are
located along the fifth dimension. We construct the most general model with above 
properties, 
calculate the 4-dimensional Planck scale and GUT scale on each brane. 
It seems to us the gauge hierarchy problem can be solved easily in this scenario:
the 5-dimensional GUT scale on each brane
can be indentified as the 5-dimensional Planck scale, but,
the 4-dimensional Planck scale is generated from the low 4-dimensional
GUT scale exponentially in our world.
 The cosmology constant is sectional constant along the fifth dimension, so,
for any point in $M^4 \times R^1$ and $M^4 \times R^1/Z_2$, which is not belong to any 
brane 
 and the section with zero cosmology constant,  
there is a neighborhood which is diffeomorphic to ( or a slice of ) $AdS_5$ space.
 Furthermore, We also give two simple models which are subsets of the model with more 
branes: 
general two branes case, three branes with $Z_2$ symmetry. The gauge hierarchy problem is 
solved 
explicitly in those models. One new feature is that, sometimes, the 5-dimensional Planck 
scale might be
very large because of the exponential factor.

\section{General Solution and Model}
First, we consider the fifth dimension is $R^1$. Assuming we have $l+m+1$ parallel positive 
tension
3-branes, and
their fifth coordinastes are: $-\infty < y_{-l} < y_{-l+1} < ...< y_{-1} < y_0
< y_{1} < ... < y_{m-1} < y_{m} < +\infty$.  
The 5-dimensional metric in these  branes are:
\begin{equation}
\label{smmetric}
g_{\mu \nu}^{(i)} (x^{\mu}) \equiv G_{\mu \nu}(x^{\mu}, y=y_i) ~,~\,
\end{equation}
where $G_{AB}$ is the five-dimensional metric, and $A, B = \mu, y$~\footnote{
We assume that $G_{ \mu 5} = 0 $ here. In addition, if the fifth dimesnion has
$Z_2$ symmetry, i. e.,  the Lagrangian is invariant under the 
transformation $ y \leftrightarrow -y$, then, $G_{ \mu 5} = 0 $.}.

The classical Lagrangian is given by:

\begin{eqnarray}
S &=& S_{gravity} + S_B 
~,~\,
\end{eqnarray}
\begin{eqnarray}
S_{gravity} &=& 
\int d^4 x  ~dy~ \sqrt{-G} \{- \Lambda (y) + 
{1\over 2} M_X^3 R \} 
~,~\,
\end{eqnarray}
\begin{eqnarray}
S_B &=& \sum_{i=-l}^{m} \int d^4 x \sqrt{-g^{(i)}} \{ {\cal L}_{i} 
-  V_{i} \} 
~,~\,
\end{eqnarray}
where $M_X$ is the 5-dimensional Planck scale,
$\Lambda (y)$ is the cosmology constant, and 
$V_i$ where $i=-l, ..., m$ is the brane tension.
The $\Lambda(y)$ is defined as the following:
\begin{eqnarray}
\Lambda (y) &=& \sum_{i=1}^m \Lambda_i \left(\theta (y-y_{i-1}) - \theta (y-y_i) \right)
+ \Lambda_{+\infty} \theta (y-y_m)
\nonumber\\&& +
\sum_{i=-l+1}^0 \Lambda_i \left(\theta (-y+y_i) - \theta (-y+y_{i-1}) \right)
+ \Lambda_{-\infty} \theta ( -y + y_{-l} ) 
~,~\,
\end{eqnarray}
where $\theta (x) = 1$ for $x \geq 0$ and $\theta (x) = 0$ for $x < 0$. So,
$\Lambda(y) $ is sectional constant.

The 5-dimensional Einstein equation for the  above action is:
\begin{eqnarray} 
\sqrt{-G} \left( R_{AB}-{1 \over 2 } G_{AB} R \right) &=& - \frac{1}{ M_X^3} 
[ \Lambda (y) \sqrt{-G} ~G_{AB} +  
\nonumber\\&& \sum_{i=-l}^{m}
V_{i} \sqrt{-g^{(i)}} ~g_{\mu \nu}^{(i)} 
~\delta^\mu_M \delta^\nu_N ~\delta(y-y_i)  ] ~.~ \,
\end{eqnarray}
Assuming that there exists a solution that
 respects   4-dimensional 
Poincare invariance in the $x^{\mu}$-directions, one obtains
the 5-dimensional metric:
\begin{eqnarray} 
ds^2 = e^{- 2 \sigma(y)} \eta_{\mu \nu} dx^{\mu} dx^{\nu}
 + dy^2 ~.~\, 
\end{eqnarray}
With this metric, the Einstein equation reduces to:
\begin{eqnarray}
\sigma^{\prime 2} = { - {\Lambda (y) } \over { 6 M_X^3}} ~,~
 \sigma^{\prime \prime} =  \sum_{i=-l}^{m}
{{V_{i}} \over\displaystyle {3 M_X^3 } } \delta (y-y_i)
~.~\, 
\end{eqnarray}

The general solution to  above differential equations is:
\begin{equation}
\sigma (y) = \sum_{i=-l}^m k_i |y-y_i| + k_c y + c 
~,~\,
\end{equation}
where $k_c$ and c are constants, and $k_i > 0$ for $i=-l, ..., m$.
The relations between the $k_i$ and $V_i$, 
and the relations between the $k_i$  and $\Lambda_i$ are:
\begin{equation}
V_i= 6 k_i M_X^3
~,~
\Lambda_i= -6 M_X^3 (\sum_{j=i}^m k_j - \sum_{j=-l}^{i-1} k_j-k_c)^2
~,~\,
\end{equation}
\begin{equation}
\Lambda_{-\infty}= -6 M_X^3 (\sum_{j=-l}^m k_j-k_c)^2
~,~ \Lambda_{+\infty}= -6 M_X^3 (\sum_{j=-l}^m k_j+ k_c)^2
~.~\,
\end{equation}
Therefore, the cosmology constant is negative except the section (at most one)
with zero cosmology constant, then, for any point in 
$M^4 \times R^1$, 
 which is not belong to any brane and the section with zero cosmology constant, 
there is a neighborhood which is diffeomorphic to ( or a slice of ) $AdS_5$ space.
Moreover, the
cosmology constant and brane tensions should satisfy above equations. In order to obtain 
finite
4-dimensional Planck scale, we obtain the constraints: $ \sum_{j=-l}^m k_j > | k_c |$.

The general bulk metric is:
\begin{equation}
ds^2 = e^{-2 \sum_{i=-l}^m k_i |y-y_i| -2 k_c y - 2 c} \eta_{\mu \nu} dx^{\mu} dx^{\nu} + 
dy^2
~.~ \,
\end{equation}

And the corresponding 4-dimensional Planck scale  is:
\begin{equation}
M_{pl}^2 =  M_X^3 \left( T_{-\infty, -l} + T_{m, +\infty} + \sum_{i=-l}^{m-1} T_{i, i+1} 
\right) 
 ~,~ \,
\end{equation}
where 
\begin{equation}
T_{-\infty, -l} = {1 \over\displaystyle {2 \chi_{-\infty}}} e^{-2 \sigma (y_{-l})}
 ~,~ 
T_{m, +\infty} = {1 \over\displaystyle {2 \chi_{+\infty}}} e^{-2 \sigma (y_{m})}
 ~,~ \,
\end{equation}
if $\chi_{i, i+1} \neq 0$, then
\begin{equation}
T_{i, i+1} = {1 \over\displaystyle {2 \chi_{i, i+1}}} \left( e^{-2 \sigma (y_{i+1})}
 - e^{-2 \sigma (y_i)} \right)
 ~,~ \,
\end{equation}
and if $\chi_{i, i+1} =0$, then
\begin{equation}
T_{i, i+1} = (y_{i+1}- y_i) e^{-2  \sigma (y_i)}
 ~,~ \,
\end{equation}
where
\begin{equation}
\chi_{\pm \infty} = \sum_{j=-l}^m k_j \pm k_c ~,~ 
\chi_{i, i+1} =
\sum_{j=i+1}^m k_j - \sum_{j=-l}^i k_j - k_c
 ~.~ \,
\end{equation}
By the way, one can easily prove that $T_{i, i+1}$ is positive, which makes sure that
the 4-dimensional Planck scale is positive.

In addition, the four dimensional GUT scale on i-th brane $M_{GUT}^{(i)}$ is
related to the five dimensional GUT scale on i-th brane $M5_{GUT}^{(i)}$:
\begin{equation}
M_{GUT}^{(i)} = M5_{GUT}^{(i)} e^{-\sigma (y_i)} 
 ~.~ \,
\end{equation}
In this paper, we assume that $ M5_{GUT}^{(i)} \equiv M_X $, for  $i =-l, ..., m$.

This solution can be generalized to the solution 
with $Z_2$ symmetry. Because of $Z_2$ symmetry, $k_c = 0$.
 There are two kinds of such models, one is the odd number
of the branes, the other is the even number of the branes. For the first one,
one just requires that $k_{-i} = k_i$, $ y_{-i} = - y_i$,
 and $m=l$. For the second case,
one just requires that $k_{-i} = k_i$, $y_{-i} = -y_i$,
 $m=l$, and $k_0=0$ (no number 0 brane). Furthermore, the solution
 can also be generalized to the case in which the fifth dimension is $R^1/Z_2$,
 one just requires that $k_{-i} = k_i$, $ y_{-i} = - y_i$,
 $m=l$, then, introduces the equivalence classes: $ y \sim - y$
 and $i-th ~brane \sim  (-i)-th ~brane$. The only  trick point in 
 this case is that the brane tension $V_0$ is half of the original value, i. e., 
$V_0= 3 k_0 M_X^3$.

\section{Two Simple Models} 
Randall and Sundrum constructed the model with only one 3-brane which is the simplest model 
in above general
construction~\cite{LRRSN}.
 However, the gauge hierarchy problem is not solved in that model.
In this section, we will give two explicit simple models without gauge hierarchy problem,
because all the other models with more branes will have these simple models as subset.

(I) Two brane cases with $k_c$: their positions are $y_0$, $y_1$, respectively, and the 
values of the brane 
tension divided by $6 M_X ^3$ are: $ k_0 $, $k_1$, respectively. 
And the constraint is $ k_0 + k_1 > |k_c|$. Therefore, we obtain:
\begin{equation}
\sigma (y) = k_0 | y-y_0 | + k_1 | y- y_1| + k_c y + c 
~,~\,
\end{equation}
\begin{equation}
\sigma (y_0) = k_1 ( y_1 -y_0 ) + k_c y_0 + c
~,~
\sigma (y_1) = k_0 ( y_1 - y_0) + k_c y_1 + c
~.~\,
\end{equation}

Assuming no section has zero cosmology constant, one obtains the four dimensional Planck 
scale:
\begin{equation}
M_{pl}^2 = {M_X^3}  \left({{ k_0} \over\displaystyle { k_0^2 - (k_1-k_c)^2}} e^{-2 
\sigma(y_0)} + 
{{ k_1} \over\displaystyle { k_1^2 - (k_0+ k_c)^2}} e^{-2 \sigma(y_1)} \right)
~.~\,
\end{equation}

Without loss of generality, assuming that $ \sigma(y_0) < \sigma(y_1) $
and $e^{-2 (\sigma(y_1)-\sigma(y_0))} << 1$, we obtain
\begin{equation}
M_{pl}^2 = {M_X^3} {{ k_0} \over\displaystyle { k_0^2 - (k_1-k_c)^2}} e^{-2 \sigma(y_0)}
~.~\,
\end{equation}
If the brane with position $y_1$ is the observable brane, assuming
 $M_X = { { k_0^2 - (k_1-k_c)^2} \over\displaystyle {k_0}}$, we obtain:
\begin{equation}
M_{GUT}^{(1)} = M_{pl} e^{-(\sigma(y_1) - \sigma(y_0)) } 
 ~.~ \,
\end{equation}
So, we can push the GUT scale in our world to TeV scale and
$10^5$ GeV scale range if  $(k_0 -k_1+k_c) ( y_1 -y_0 )$ = 34.5 and 30, 
respectively. And the value of $\sigma(y_0)$ determines the
relation between $M_{pl}$ and $M_X$.
Explicit example: $k_0=k_1=k_c > 0$, $ k_c( y_1 -y_0 )$ = 34.5 and 30. 
So, $k_c$ is also an important factor to solve the gauge hierarchy problem. 

If the brane with position $y_0$  is the observable brane,
 we can solve the gauge hierarchy problem only when $ \sigma(y_0) > 0$. Assuming that
$M_{pl} = { { k_0^2 - (k_1-k_c)^2} \over\displaystyle {k_0}} $ and  
$e^{-2 (\sigma(y_1)-\sigma(y_0))} < < 1$, we obtain:
\begin{equation}
M_{GUT}^{(0)} = M_{pl} e^{-{1\over 3} \sigma(y_0)  } 
 ~,~ \,
\end{equation}
with $\sigma(y_0)$ = 103.5 and 90, we can have GUT scale in
our world at TeV scale and $10^5$ GeV scale, respectively.
The five-dimensional Planck scale is
$10^{48}$ GeV and $10^{44}$ GeV, respectively. 

(II) Three 3-branes with $Z_2$ symmetry,
their positions are: $-y_1$, $0$, $y_1$, respectively, and their
brane tensions divided by $6 M_X^3$ are: $k_1, k_0, k_1$, respectively. 
One  obtains $M_{pl}$, and the 4-dimensional GUT scales
on the branes at y=0 and at y=$y_1$:
\begin{equation}
M_{pl}^2 = M_X^3  e^{-4 k_1 y_1-2c} \left({1\over\displaystyle k_0} 
- {{2 k_1} \over\displaystyle {(2 k_1+k_0) k_0}}e^{-2 k_0 y_1} \right) 
~,~\,
\end{equation}
\begin{equation}
M_{GUT}^{(0)} =  M5_{GUT}^{(0)} e^{-2 k_1 y_1-c}
 ~,~ 
M_{GUT}^{(1)} =  M5_{GUT}^{(1)} e^{-2 k_1 y_1-k_0 y_1-c}
 ~,~ \,
\end{equation}   
where $ M5_{GUT}^{(0)} = M5_{GUT}^{(1)} = M_X$.
If the  brane  with position $y_1$  is the observable brane,
one can sovle the gauge hierarchy problem. For example,
assuming $e^{-2 k_0 y_1} << 1$ and $ k_0 =M_X$, one obtains:
\begin{equation}
M_{pl}=M_X e^{-2 k_1 y_1-c}
 ~,~ 
M_{GUT}^{(1)} = M_{pl}  e^{-k_0 y_1}
 ~.~ \,
\end{equation}
With $k_0 y_1$ = 34.5 and 30, one obtains the GUT scale in
our world will be at TeV scale and $10^5$ GeV scale range, respectively.
And the relation between $M_{pl}$ and $M_X$ depends on the
value of $2 k_1 y_1 + c$.  In addition,
if  the  brane with position $y=0$ is the observable brane,
  the gauge hierarchy problem can be solved, too.
Assuming $M_{pl}= k_0$ and $e^{-2k_0 y_1}$ is very small, one obtains:
\begin{equation}
M_{pl}=  M_X e^{-{4\over 3} k_1 y_1- {{2}\over 3} c}
 ~,~ 
M_{GUT}^{(0)} = M_{pl} e^{-{2\over 3} k_1 y_1 - {1\over 3} c} 
 ~,~ \,
\end{equation}
with $ 2 k_1 y_1+c$ = 103.5 and 90, one can push the GUT scale in
our world to the TeV scale and $10^5$ GeV scale range, respectively.
And the five dimensional Planck scale is $10^{48}$ GeV and $10^{44}$ GeV, respectively.

\section*{Acknowledgments}
We would like to thank M. Cvetic, K. R. Dienes, C. Grojean, H. D. Kim and
 T. Shiromizu for bringing my attention to their papers and helpful comments.
This research was supported in part by the U.S.~Department of Energy under
 Grant No.~DE-FG02-95ER40896 and in part by the University of Wisconsin 
 Research Committee with funds granted by the Wisconsin Alumni
  Research Foundation.

\end{document}